# Parallel and Intelligent Bayesian Optimization for PCB Stack-up Design

Jiayi He[#1], Aravind Sampath Kumar[$2], Arun Chada[*3], Bhyrav Mutnury[*4], James Drewniak[#5]

[#]EMC Laboratory, Department of Electrical and Computer Engineering, Missouri University of Science and Technology, Rolla, Mo, USA
[1]hejiay@mst.edu, [5]drewniak@mst.edu

[*]Dell, Enterprise Product Group, One Dell Way, MS RR5-31, Round Rock, Texas, USA
[3]arun_chada@dell.com, [4]bhyrav_mutnury@dell.com

[$]Department of Electrical and Computer Engineering, North Carolina State University, Raleigh, NC, USA
[2]asampat2@ncsu.edu

*Abstract*—**Design of printed circuit board (PCB) stack-up requires the consideration of characteristic impedance, insertion loss and crosstalk. As there are many parameters in a PCB stack-up design, the optimization of these parameters needs to be efficient and accurate. A less optimal stack-up would lead to expensive PCB material choices in high speed designs. In this paper, an efficient global optimization method using parallel and intelligent Bayesian optimization is proposed for the stripline design.**

## I. Introduction

In high speed system design, optimizing printed circuit board (PCB) stack-up is playing a more and more important role in design stage. Since PCB stack-up definition is one of the first things to be locked during the design phase, a less than optimum PCB stack-up can result in the selection of expensive laminate materials. The cost of a system can be significantly reduced with an optimum stack-up design.

The design of PCB stack-up usually considers the transmission line electrical properties, such as impedance matching, insertion loss minimization and crosstalk minimization. There are many tools available for designers to get these transmission line properties by inputting the stack-up parameters. However, optimizing these parameters is not easy as there are many parameters at play. For a differential stripline shown in Figure 1, the input parameters include the trace width (W), trace spacing (S), trace thickness (T), core height ($H_1$), total dielectric height (H), dielectric constant for core and prepreg ($\varepsilon_r$). Designer should also consider other parameters like loss tangent, surface roughness and etching factor to name a few. There are 6-16 parameters that need to be optimized depending on the engineer's requirement.

Optimization of the stack-up needs running the transmission line solver and brute-force, full-factorial search of the design space can take weeks and months of simulation time as the number of combinations would run into 100s of thousands. A smarter method is needed to reduce the search space from 100s of thousands to few hundreds. In this paper, Bayesian optimization (BO) is used to perform the global optimization in order to reduce the number of objective function evaluations. However, classic BO also has its limitation in convergence speed when the dimension of the objective function or the number of observations gets large. In this paper, a parallel and intelligent BO is proposed for the PCB stack-up optimization.

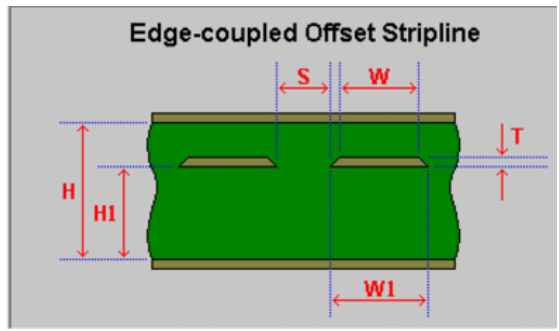

Fig. 1 A sample cross-section of a PCB

The paper is organized as follows: Section II describes Bayesian optimization. The limitations of Bayesian optimization are discussed in Section III. Section IV discussed the parallel intelligent Bayesian optimization along with the simulation results. Section V summarizes the paper.

## II. Bayesian Optimization with Gaussian Process

Bayesian Optimization consist of two main components: a Bayesian statistics model for modeling the objective function and an acquisition function for determining which point to sample next. Usually Gaussian process (GP) regression is used as the Bayesian statistical model. For GP priors, a joint Gaussian distribution is created with the entire set of available observation points. The objective function "*f*" is defined as a GP prior with a mean function "*μ*" and a covariance function "*k*." Based on prior observation points, the prior distribution on [$f(x_1), …, f(x_N)$] is:

$$f(x_{1:N}) \sim N(\mu(x_{1:N}), K) \tag{1}$$

where $x_{1:N}$ represent the "*N*" observation points, $\mu(x_{1:N})$ is the mean vector at the corresponding observation points and *k* is the corresponding covariance matrix:

$$K = \begin{bmatrix} k(x_1, x_1) & \cdots & k(x_1, x_N) \\ \vdots & \ddots & \vdots \\ k(x_N, x_1) & \cdots & k(x_N, x_N) \end{bmatrix} \tag{2}$$

Where the kernel function *k* is defined by:

$$k(x_i, x_j) = \left(1 + \frac{\sqrt{5}r}{\theta} + \frac{5r^2}{3\theta^2}\right)\exp\left(-\frac{\sqrt{5}r}{\theta}\right) \quad (3)$$

$$r = ||x_i - x_j|| \quad (4)$$

where θ is a hyperparameter for effective length scaling.

To infer the value of *f(*x*)* at the next data point, $x_{N+1}$, we can compute the conditional distribution of *f(x)* given these observations using Bayes' theorem.

$$f(x_{N+1})|f(x_{1:N}) \sim N(\mu(x_{N+1}), \sigma^2(x_{N+1})) \quad (5)$$
$$\mu(x_{N+1}) = \boldsymbol{k}^T K^{-1} f(x_{1:N}) \quad (6)$$
$$\sigma^2(x_{N+1}) = k(x_{N+1}, x_{N+1}) - \boldsymbol{k}^T K^{-1} \boldsymbol{k} \quad (7)$$
$$\boldsymbol{k} = [k(x_{n+1}, x_1) \quad \cdots \quad k(x_{n+1}, x_N)]^T \quad (8)$$

Where K is the covariance matrix and *k* is the kernel function. Then the posterior distribution of any new data point is given by the above equations.

The next point to be evaluated is chosen by maximizing or minimizing the acquisition function. There are three widely used acquisition function: probability of improvement (PI), expected improvement (EI) and upper/lower confidence bound (UCB/LCB). The goal of acquisition function is to find the point that potentially improves the current best or worst value. In this paper, PI is used to minimize the objective function. The next sample is given by:

$$x_{N+1} = \text{argmin}(\mu(x_i) - \tau\sigma(x_i)) \quad (9)$$

Where $\mu(x_i)$ and $\sigma(x_i)$ are the posterior distribution calculated from (6) and (7), τ is a hyperparameter that determines the ratio of exploration and exploitation. Larger τ leads to more exploration than exploitation.

A typical flow chart of BO is shown in Figure 2:

1) Choose N initial samples and evaluate the objective function.
2) Train the GP regressor from the observed points and calculate the posterior distribution.
3) Add a new sample based on the acquisition function and evaluate this sample.
4) Repeat step 2 and 3 until the stopping criteria is met.

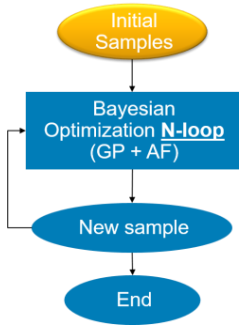

Fig. 2. Flow chart of typical BO

In BO process, since the acquisition function is only determined by previous observations, a new sample is selected without computing the objective function in advance and only the selected sample point will be evaluated. Thus, the number of samples that are evaluated is minimized in BO. As evaluating the objective function at one sample means running the transmission line solver once, the computation time can be significantly reduced by minimizing the number of computation samples.

### III. CHALLENGES OF BAYESIAN OPTIMIZATION

The main challenges of Bayesian optimization are its feasibility to scale to higher dimensions and its efficiency at large number of observations. When the number of observations gets larger, the size of the covariance matrix also gets larger, resulting in a significant increase in computation time. Higher dimensional problems force BO to have a larger initial sampling or running for more iterations, both will cause BO take longer time to converge.

Another challenge is the optimized result may be affected by the initial samples. Since the GP posterior is determined by previous observations, the initial random samples and its size will play a role in it. The optimized result and the number of iteration needed may vary from run to run. This phenomenon is more obvious when the dimension of the problem gets larger, as BO forces to have a larger initial sample size at higher dimensions.

The choice of covariance function and acquisition function will also influence the efficiency of BO. There are many different covariance functions such as the squared exponential function, the Matern function, the rational quadratic function, etc. The choice of covariance function may affect the convergence speed of BO. In this paper, we used the Matern covariance function. In LCB acquisition function, the hyperparameter, τ, represents the factor between exploration and exploitation, so the value of τ needs to be tuned to achieve the global optimum with fewer iterations.

### IV. PARALLEL AND INTELLIGENT BAYESIAN OPTIMIZATION

To resolve the challenges in classic BO, a parallel and intelligent Bayesian optimization method is proposed. The flow chart of this method is shown in Figure 3. In first step of this algorithm, multiple small BOs are run in parallel with different random initial samples. Each BO follows the procedure of classical Bayesian optimization as described in Section III. After "N" iterations, the data collected by all individual BOs are combined to a bigger dataset. Then in second step, another large BO is performed on the combined dataset for "n" more iterations and reports the optimized result.

In the proposed method, multiple independent small BOs are running parallelly. The size of initial samples is relative small, so the computation speed at this step is fast. When the dataset gets combined, the calculation speed will be slower as the number of observations gets large, but we can limit the number of iterations for this BO as previous steps already give results very close to the global optimum. So the computation time challenge at larger dataset gets resolved.

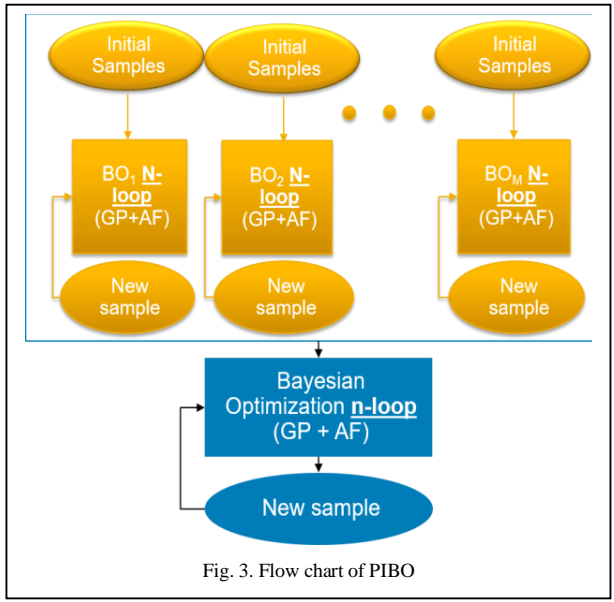

Fig. 3. Flow chart of PIBO

This method also mitigates the effect of random initial samples. By running multiple BOs with different initial samples and combining the data from each BO, the randomness in initial data is unlikely to play a role in this procedure any longer. This feature makes the algorithm more robust.

Considering the fact that the choice of hyperparameters will affect the efficiency of almost every machine learning algorithm, it is acceptable that the covariance function and hyperparameters in the acquisition function still play a role in the proposed method. Since our method is more efficient and not sensitive to the randomness, tuning the hyperparameters would be faster and easier.

To validate our proposed PIBO method, it is applied to a differential stripline design problem. In this preliminary study, the number of parameters are equal to 6 for simplicity. The parameters to be optimized included the trace width (W), trace spacing (S), trace thickness (T), core height ($H_1$), total height ($H_2$) and the dielectric constant ($\varepsilon_r$). The range and step of each input variables is listed in Table. 1. The total number of combinations are roughly 100K.

TABLE I. RANGE AND STEP OF EACH PARAMETER

| Parameter | Min | Max | Step |
|---|---|---|---|
| W (mil) | 3 | 8 | 0.25 |
| S (mil) | 3 | 8 | 0.25 |
| T (mil) | 3 | 5 | 0.5 |
| $H_1$ (mil) | 8 | 10 | 0.5 |
| $H_2$ (mil) | 1.1 | 1.3 | 0.1 |
| $\varepsilon_r$ | 3.6 | 3.8 | 0.1 |

The goal of this optimization is to minimize the insertion loss while the characteristic impedance matches the target impedance, which is 85Ω. The characteristic impedance and the insertion loss are calculated from a 2D cross-sectional solver. The objective function is defined as:

$$f = |Z_c - Z_T| + 100 * \text{loss} \quad (10)$$

Where $Z_c$ and loss are the characteristic impedance and the loss of the transmission line, $Z_T$ is the target impedance. The goal is to minimize this objective function. The number 100 is a factor to adjust the weight of these two terms, which is tuned by some experiments. If the designer aims to maximize the loss while matching the target impedance, the loss term can be replaced by the inverse of the loss and the factor needs to be tuned again as shown below:

$$f = |Z_c - Z_T| + 40/\text{loss} \quad (11)$$

Using the proposed method, the optimum design is obtained as: 7.25mils for trace width, 7.75mils for trace spacing, 5mils for core height, 10mils for total height, 1.3mils for trace thickness and 3.6 for dielectric constant. The whole optimization process only runs 260 simulations and the computation time is 15 minutes on a 12 core machine. Running all 100K simulations on a 200 core server in parallel will take 14 hours. Compared to running all the combinations, our proposed method is very efficient. The experiment is repeated multiple times and the global optima converges around 250 to 270 simulations using the proposed approach

## V. CONCLUSION

In this paper, a parallel and intelligent Bayesian optimization method is proposed. By running BO in parallel, the computation time gets reduced and the repeatability is improved. This method is successfully applied to PCB stack-up design to minimize the loss of transmission line and match the target impedance. In the future, this method will be applied to more complex problems with higher dimension and more outputs (like crosstalk). An approach to find optimum initial samples and hyperparameters will also be studied.